# A family of models for spherical stellar systems

SCOTT TREMAINE
Canadian Institute for Theoretical Astrophysics, University of Toronto
60 St. George St., Toronto M5S 1A7, Canada
Electronic mail: tremaine@cita.utoronto.ca

DOUGLAS O. RICHSTONE
Dept. of Astronomy, University of Michigan, Ann Arbor, MI 48109
and Institute for Theoretical Physics, University of California
Santa Barbara, CA 93106-4030
Electronic mail: dor@astro.lsa.umich.edu

YONG-IK BYUN
Institute for Astronomy, University of Hawaii
2680 Woodlawn Dr., Honolulu, HI 96822
Electronic mail: byun@haneul.ifa.hawaii.edu

ALAN DRESSLER
Carnegie Observatories, 813 Santa Barbara St., Pasadena, CA 91101
Electronic mail: dressler%ociw1@caltech.edu

S. M. FABER
UCO/Lick Observatories
Board of Studies in Astronomy and Astrophysics
University of California, Santa Cruz, CA 95064
Electronic mail: faber@helios.ucsc.edu

CARL GRILLMAIR
UCO/Lick Observatories
Board of Studies in Astronomy and Astrophysics
University of California, Santa Cruz, CA 95064
Electronic mail: carl@helios.ucsc.edu

JOHN KORMENDY
Institute for Astronomy, University of Hawaii
2680 Woodlawn Dr., Honolulu, HI 96822
Electronic mail: kormendy@oort.ifa.hawaii.edu

TOD R. LAUER
Kitt Peak National Observatory
National Optical Astronomy Observatories, Tucson, AZ 85719
Electronic mail: lauer@noao.edu

ABSTRACT

We describe a one-parameter family of models of stable spherical stellar systems in which the phase-space distribution function depends only on energy. The models have similar density profiles in their outer parts ($\rho \propto r^{-4}$) and central power-law density cusps, $\rho \propto r^{3-\eta}$, $0 < \eta \leq 3$. The family contains the Jaffe (1983) and Hernquist (1990) models as special cases. We evaluate the surface brightness profile, the line-of-sight velocity dispersion profile, and the distribution function, and discuss analogs of King's core-fitting formula for determining mass-to-light ratio. We also generalize the models to a two-parameter family, in which the galaxy contains a central black hole; the second parameter is the mass of the black hole. Our models can be used to estimate the detectability of central black holes and the velocity-dispersion profiles of galaxies that contain central cusps, with or without a central black hole.

# 1 Introduction

Despite the increasing sophistication of numerical models of stellar systems, analytic or semi-analytic models still have many uses: they offer insight into the behavior of more realistic and complicated models, they provide tests of and initial conditions for $N$-body experiments, they are free of noise and spurious numerical relaxation, and in some cases they fit observed stellar systems remarkably well.

An additional motivation for the models described below is recent high-resolution ground-based and Hubble Space Telescope photometry of elliptical galaxies and bulges. These observations typically fail to resolve the central regions into constant density cores, even at a resolution of $0''.1$. Specific examples include M32 and M87 (Lauer et al. 1992a,b). HST observations of a sample of more than 30 elliptical and S0 galaxies by our group (Lauer et al. 1993) establish that, at a resolution of $< 0''.1$, the surface brightness profile $I(R)$ is usually best approximated by $R^{-\gamma}$, with $0 < \gamma \lesssim 1$. Most previous dynamical studies of galaxies have focused on models having a constant density core, that is, an analytic surface brightness that can be expanded as a Taylor series in the form $I(R) \propto 1 - AR^2 + \ldots$. Thus the HST observations demand new dynamical models to predict the kinematic properties of the central regions and to assess whether massive central black holes are implied by the observations. A natural first step is to seek appropriate analytic models.

Among the simplest stellar systems are isotropic spherical systems, in which the gravitational potential $\Phi$ and the density $\rho$ are spherically symmetric and the phase-space distribution function depends only on the energy per unit mass, $f = f(E)$ where $E = \frac{1}{2}v^2 + \Phi$ (e.g. Binney and Tremaine 1987, hereafter BT). We shall focus on the case in which the mass-to-light ratio $\Upsilon$ is independent of radius, so that the emissivity is $\rho(r)/\Upsilon$.

Two of the most successful analytic models for elliptical galaxies and the bulges of disk galaxies are the Jaffe (1983) model,

$$\rho(r) = \frac{1}{4\pi r^2 (1+r)^2}, \tag{1}$$

and the Hernquist (1990) model,

$$\rho(r) = \frac{1}{2\pi r (1+r)^3}. \tag{2}$$

Both models are normalized so that the total mass $M = 4\pi \int \rho(r) r^2 dr$ is unity. The Jaffe and Hernquist models have similar profiles in their outer parts ($\rho \propto r^{-4}$); both also have central density cusps, but they differ in the strength of the cusp ($\rho \propto r^{-2}$ and $r^{-1}$ respectively).

In this paper we describe a one-parameter family of isotropic spherical stellar systems, which we call $\eta$-models, that preserves most of the analytic simplicity of the Jaffe and Hernquist models, and which includes them as special cases. The parameter $\eta$ describes the strength of the central cusp, which diverges as $\rho \propto r^{\eta-3}$ for $0 < \eta \leq 3$. We describe these models in §2. In §3 we generalize the models to a two-parameter family containing a non-luminous central mass ("black hole" for short); the second parameter is the mass of the black hole. Sections 4 and 5 contain discussion.

The work described in this paper has been anticipated to varying extents by several authors. Hernquist (1990, eq. 43) briefly describes a three-parameter family that includes the $\eta$-models, but does not discuss the properties of any systems except the Jaffe and Hernquist models. Saha (1993) writes down the potential and density distributions of the $\eta$-models, but focuses on their use as basis functions for solving Poisson's equation rather than as models of stellar systems. As this paper neared completion, we received a preprint from Dehnen (1993) that independently derives many of the properties of $\eta$-models discussed in §2, but does not discuss models with a central black hole.

## 2 Models

We choose units in which the total mass $M$ of the stars and the gravitational constant $G$ are both unity.

We define a family of spherical stellar systems called "$\eta$–models" by the density distribution

$$\rho_\eta(r) \equiv \frac{\eta}{4\pi} \frac{1}{r^{3-\eta}(1+r)^{1+\eta}}, \qquad 0 < \eta \leq 3. \tag{3}$$

We must have $\eta > 0$[1]; we show below that models with $\eta > 3$ (density approaching zero at the origin) are unphysical. All $\eta$ models have $\rho \propto r^{-4}$ as $r \to \infty$. Jaffe's model (eq. 1) corresponds to $\eta = 1$ and Hernquist's model (eq. 2) corresponds to $\eta = 2$. The density diverges near the origin for all models with $\eta < 3$; only models with $\eta = 3$ have constant-density cores.

---

[1] Saha (1993) points out that the model with $\eta = 0$ has a potential corresponding to a unit point mass (see eq. 5 below). Thus it could represent a stellar system with luminosity density $\propto 1/[r^3(1+r)]$, negligible mass-to-light ratio, and a central black hole.

**Mass distribution**    The mass interior to radius $r$ is

$$M_\eta(r) \equiv 4\pi \int_0^r r^2 \rho_\eta(r) dr = \frac{r^\eta}{(1+r)^\eta}. \tag{4}$$

**Potential**    The gravitational potential is

$$\begin{aligned}\Phi_\eta(r) &\equiv -\int_r^\infty \frac{M_\eta(r)dr}{r^2} = \frac{1}{\eta-1}\left[\frac{r^{\eta-1}}{(1+r)^{\eta-1}} - 1\right], & \eta \neq 1, \\ &= -\ln(1+1/r), & \eta = 1,\end{aligned} \tag{5}$$

where we assume $\Phi(r) \to 0$ as $r \to \infty$. The depth of the potential well near the origin is finite for $\eta > 1$, $\Phi_\eta(0) = -(\eta-1)^{-1}$, while for $0 < \eta < 1$ the potential near the origin diverges, $\Phi_\eta(r) \to -(1-\eta)^{-1} r^{-(1-\eta)}$.

**Velocity dispersion**    We assume that the phase-space distribution function (DF) depends only on energy, which implies that the velocity-dispersion tensor is isotropic. Thus the radial velocity dispersion $\overline{v_r^2}^{1/2}(r)$ satisfies the hydrostatic equilibrium equation,

$$\frac{d(\rho \overline{v_r^2})}{dr} = -\rho \frac{d\Phi}{dr}, \tag{6}$$

with boundary condition $\rho \overline{v_r^2} = 0$ as $r \to \infty$. Since $\rho$ and $d\Phi/dr$ are non-negative, $\overline{v_r^2}(r)$ is always positive. Thus we have

$$\begin{aligned}\overline{v_{r,\eta}^2}(r) &= \frac{1}{\rho_\eta(r)} \int_r^\infty \frac{M_\eta(r)\rho_\eta(r)}{r^2} dr \\ &= r^{3-\eta}(1+r)^{1+\eta} \left(\frac{1}{2\eta-4} - \frac{4}{2\eta-3} + \frac{3}{\eta-1} - \frac{4}{2\eta-1} + \frac{1}{2\eta}\right) \\ &\quad - \frac{r^{\eta-1}(1+r)^{5-\eta}}{2\eta-4} + \frac{4r^\eta(1+r)^{4-\eta}}{2\eta-3} - \frac{3r^{\eta+1}(1+r)^{3-\eta}}{\eta-1} \\ &\quad + \frac{4r^{\eta+2}(1+r)^{2-\eta}}{2\eta-1} - \frac{r^{\eta+3}(1+r)^{1-\eta}}{2\eta}.\end{aligned} \tag{7}$$

This result is not valid for the four cases $\eta = \frac{1}{2}, 1, \frac{3}{2}, 2$:

$$\begin{aligned}\overline{v_{r,1/2}^2}(r) &= \tfrac{1}{3} r^{-1/2}(1+r)^{9/2} - 2r^{1/2}(1+r)^{7/2} + 6r^{3/2}(1+r)^{5/2} \\ &\quad - r^{5/2}(1+r)^{3/2}\left[4\ln(1+1/r) + \tfrac{10}{3}\right] - r^{7/2}(1+r)^{1/2}, \\ \overline{v_{r,1}^2}(r) &= \tfrac{1}{2} - 2r - 9r^2 - 6r^3 + 6r^2(1+r)^2 \ln(1+1/r), \\ \overline{v_{r,3/2}^2}(r) &= r^{1/2}(1+r)^{7/2} - r^{3/2}(1+r)^{5/2}\left[4\ln(1+1/r) - \tfrac{10}{3}\right] \\ &\quad - 6r^{5/2}(1+r)^{3/2} + 2r^{7/2}(1+r)^{1/2} - \tfrac{1}{3}r^{9/2}(1+r)^{-1/2}, \\ \overline{v_{r,2}^2}(r) &= r(1+r)^3 \left[\ln(1+1/r) - \tfrac{25}{12}\right] + 4r^2(1+r)^2 - 3r^3(1+r) + \tfrac{4}{3}r^4 - \tfrac{1}{4}r^5(1+r)^{-1}.\end{aligned} \tag{8}$$

These formulae can be difficult to evaluate at large $r$ because of near-cancellations between the terms. An alternate expression is the asymptotic series

$$\overline{v_{r,\eta}^2}(r) = r^{3-\eta}(1+r)^{1+\eta} \sum_{j=0}^{\infty} \frac{(-2\eta-1)\cdots(-2\eta-j)}{j!} \frac{1}{(5+j)r^{5+j}}. \quad (9)$$

Thus as $r \to \infty$, $\overline{v_r^2}(r) \to \frac{1}{5}r^{-1}$ for all $\eta$.

As $r \to 0$,

$$\overline{v_{r,\eta}^2}(r) \to \begin{cases} r^{\eta-1}/(4-2\eta), & \text{for } 0 < \eta < 2, \\ r(-\ln r - \frac{25}{12}), & \text{for } \eta = 2, \\ r^{3-\eta}\left[\frac{1}{2\eta-4} - \frac{4}{2\eta-3} + \frac{3}{\eta-1} - \frac{4}{2\eta-1} + \frac{1}{2\eta}\right], & \text{for } 2 < \eta \leq 3. \end{cases} \quad (10)$$

Notice the interesting range of behavior: the mean-square radial velocity diverges as $r^{-1}$ when $\eta$ is near zero, then diverges less and less rapidly as $\eta$ increases towards 1. Once $\eta > 1$, the mean-square velocity converges to zero, converging more and more rapidly until $\eta = 2$, when it converges as $r$. Above $\eta = 2$ the mean-square velocity converges to zero less and less rapidly, until at $\eta = 3$ it is asymptotically constant. Thus the central velocity dispersion is constant and non-zero for two distinct models:

$$\overline{v_{r,\eta}^2}(r=0) = \begin{cases} \frac{1}{2} & \text{for } \eta = 1, \\ \frac{1}{30} & \text{for } \eta = 3. \end{cases} \quad (11)$$

The inner parts of the model with $\eta = 1$ resemble the singular isothermal sphere, while the inner parts of the model with $\eta = 3$ resemble the central core of the non-singular isothermal sphere or other models with constant-density cores such as King models (King 1966; BT).

The qualitative reasons for the behavior of the central dispersion are worth describing. For $\eta < 2$ the dispersion at a given radius is comparable to the circular speed $v_c$ at that radius, given by $v_c^2 = M(r)/r \to r^{\eta-1}$, which is the natural result expected from dimensional analysis of the hydrostatic equilibrium equation (replace $d/dr$ by $-1/r$ in eq. 6). The behavior of models with $\eta > 2$ is more subtle, because the mean-square velocity and density are dominated by different parts of the energy distribution: the density is dominated by low-energy stars with mean-square velocity comparable to the square of the circular speed, $\overline{v_r^2} \propto r^{\eta-1}$, while the velocity dispersion is dominated by high-energy stars with velocities of order unity. Thus the pressure $\rho\overline{v_r^2}$ is constant and order unity as $r \to 0$; since $\rho \propto r^{\eta-3}$ the mean-square velocity $\overline{v_r^2} \propto r^{3-\eta}$.

The radial velocity dispersion $\overline{v_r^2}^{1/2}$ is plotted in Figure 1 for several values of $\eta$.

**Kinetic and potential energy**   The total kinetic and potential energies are

$$\begin{aligned} T_\eta &= 6\pi \int_0^\infty r^2 \rho_\eta(r)\overline{v_{r,\eta}^2}(r)dr \\ W_\eta &= 2\pi \int_0^\infty \rho_\eta(r)\Phi_\eta(r)r^2\,dr. \end{aligned} \quad (12)$$

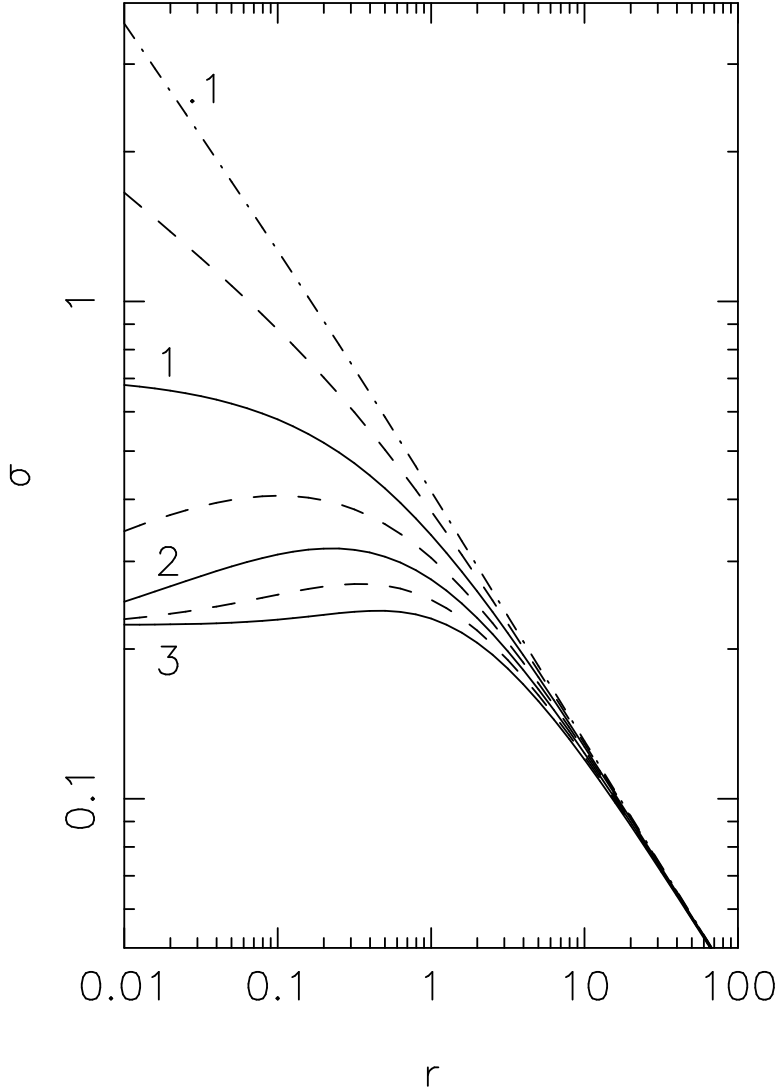

**1.** Radial velocity dispersion $\overline{v_r^2}^{1/2}$ as a function of radius (eqs. 7 and 8), for models with $\eta = 0.1$ (dash-dot curve), $\frac{1}{2}$, $\frac{3}{2}$, $\frac{5}{2}$ (unlabelled dashed curves), and 1, 2, 3 (solid curves). All models have $\overline{v_r^2}^{1/2} \to (5r)^{-1/2}$ as $r \to \infty$.

Substituting the hydrostatic equilibrium equation in the form $\rho \overline{v_r^2} = \int_r^\infty M\rho \, dr/r^2$ into the first of these equations, and the equation for the gravitational potential, $\Phi = -\int_r^\infty M \, dr/r^2$, into the second, it is straightforward to confirm the virial theorem,

$$W_\eta = -2T_\eta = -4\pi \int_0^\infty \rho_\eta(r) M_\eta(r) r \, dr. \tag{13}$$

Evaluating the integral for $\eta$-models we have

$$W_\eta = -2T_\eta = -\frac{1}{2(2\eta - 1)} \qquad \text{for } \eta > \tfrac{1}{2}; \tag{14}$$

for $\eta < \tfrac{1}{2}$ the energies are infinite.

**Distribution function**  Jeans' theorem ensures that any DF that depends only on energy is a steady-state solution of the collisionless Boltzmann equation. The density is determined by the equation

$$\rho = 2^{5/2}\pi \int_0^\psi f(\epsilon)(\psi - \epsilon)^{1/2} d\epsilon, \tag{15}$$

where the DF $f \geq 0$ is the mass per unit volume of phase space, $\epsilon \equiv -E$, and $\psi \equiv -\Phi$ (the quantities $\epsilon$ and $\psi$ are defined so that we work with non-negative variables: $\psi(r)$ is positive for all $\eta$-models at all radii, and $\epsilon$ is positive for all bound stars).

Equation (15) yields a simple proof that models with $\eta > 3$ are unphysical: since $f \geq 0$, $d\rho/d\psi = 2^{3/2}\pi \int_0^\psi f(\epsilon)(\psi - \epsilon)^{-1/2} d\epsilon$ is positive; since $d\psi/dr = -GM(r)/r^2$ is negative, $d\rho/dr = (d\rho/d\psi)/(d\psi/dr)$ must be negative. The $\eta$-models have $\rho \propto r^{\eta-3}$ near the origin, so models with $\eta > 3$ are unphysical if the DF depends only on energy.

The inverse of equation (15) is Eddington's formula (BT)

$$f(\epsilon) = \frac{1}{2^{3/2}\pi^2} \frac{d}{d\epsilon} \int_0^\epsilon \frac{d\rho}{d\psi} \frac{d\psi}{(\epsilon - \psi)^{1/2}}. \tag{16}$$

Integrating this equation by parts yields a simpler form:

$$f(\epsilon) = \frac{1}{2^{3/2}\pi^2} \int_0^\epsilon \frac{d^2\rho}{d\psi^2} \frac{d\psi}{(\epsilon - \psi)^{1/2}}, \tag{17}$$

where we have used the fact that $d\rho/d\psi = 0$ at $\psi = 0$ in $\eta$-models (as $r \to \infty$, $\psi_\eta \propto r^{-1}$ and $\rho_\eta \propto r^{-4} \propto \psi_\eta^4$). The function $d^2\rho_\eta/d\psi_\eta^2$ is most conveniently written using the intermediate parameter $u \equiv r^{-1}$

$$\frac{d^2\rho_\eta}{d\psi_\eta^2} = \frac{\eta}{4\pi} \frac{u^2[12 + 4(4-\eta)u + 2(3-\eta)u^2]}{(1+u)^{3-\eta}},$$

$$\psi_\eta = \frac{1}{\eta - 1}\left[1 - (1+u)^{1-\eta}\right], \qquad \eta \neq 1, \tag{18}$$

$$= \ln(1+u), \qquad \eta = 1.$$

It is easy to see from equations (17) and (18) that all models with $\eta \leq 3$ have $d^2\rho_\eta/d\psi_\eta^2 \geq 0$ and hence $f_\eta(\epsilon) \geq 0$, so the phase-space density is positive as required.

For $\eta > 1$ stars are present for all energies between 0 and $\epsilon_{\mathbf{max}} = 1/(\eta - 1)$; for $0 < \eta \le 1$ stars are present for all $\epsilon > 0$. The DF diverges as $\epsilon \to \epsilon_{\mathbf{max}}$ or $\epsilon \to \infty$ respectively; we have

$$\begin{aligned}
f_\eta(\epsilon) &\to \frac{\eta(3-\eta)(1-\eta)^{(1+\eta)/(1-\eta)}}{(2\pi)^{5/2}} \frac{\Gamma\left(\frac{2}{1-\eta}\right)}{\Gamma\left(\frac{1}{2} + \frac{2}{1-\eta}\right)} \epsilon^{(3+\eta)/[2(1-\eta)]}, \quad 0 < \eta < 1, \\
&\to \frac{e^{2\epsilon}}{4\pi^{5/2}}, \quad \eta = 1, \\
&\to \frac{\eta(3-\eta)(\eta-1)^{-(1+\eta)/(\eta-1)}}{(2\pi)^{5/2}} \frac{\Gamma\left(\frac{\eta+1}{\eta-1} - \frac{1}{2}\right)}{\Gamma\left(\frac{\eta+1}{\eta-1}\right)} (\epsilon_{\mathbf{max}} - \epsilon)^{-(3+\eta)/[2(\eta-1)]}, \quad 1 < \eta < 3, \\
&\to \frac{3}{4\pi^3} (\epsilon_{\mathbf{max}} - \epsilon)^{-1}, \quad \eta = 3.
\end{aligned} \tag{19}$$

At small energies,

$$f_\eta(\epsilon) \to \frac{2^{5/2}\eta}{5\pi^3} \epsilon^{5/2} \qquad \text{as } \epsilon \to 0. \tag{20}$$

Analytic expressions for the DF are available for a number of $\eta$-models:

$$\begin{aligned}
f_3(\epsilon) &= \frac{3}{4\pi^3} \left\{ 2(2\epsilon)^{1/2} \frac{3-4\epsilon}{1-2\epsilon} + 3\log\left[\frac{1-(2\epsilon)^{1/2}}{1+(2\epsilon)^{1/2}}\right] \right\}, \\
f_2(\epsilon) &= \frac{1}{2^{7/2}\pi^3(1-\epsilon)^{5/2}} \left\{ 3\sin^{-1}\epsilon^{1/2} - [\epsilon(1-\epsilon)]^{1/2}(3 + 2\epsilon - 24\epsilon^2 + 16\epsilon^3) \right\}, \\
f_{3/2}(\epsilon) &= \frac{3}{2^{9/2}\pi^3(2-\epsilon)^{9/2}} \Big\{ \tfrac{3}{2}(3 + 32\epsilon - 8\epsilon^2) \sin^{-1}(\epsilon/2)^{1/2} \\
&\quad - \frac{[\epsilon(2-\epsilon)]^{1/2}}{28}(63 + 693\epsilon - 5670\epsilon^2 + 7410\epsilon^3 - 4488\epsilon^4 + 1448\epsilon^5 - 240\epsilon^6 + 16\epsilon^7) \Big\}, \\
f_1(\epsilon) &= \frac{1}{2^{3/2}\pi^{5/2}} \left\{ -e^\epsilon \operatorname{erf}(\epsilon^{1/2}) + \frac{e^{2\epsilon}}{2^{1/2}} \operatorname{erf}[(2\epsilon)^{1/2}] - \frac{2}{\pi^{1/2}} F(\epsilon^{1/2}) + \frac{2^{1/2}}{\pi^{1/2}} F[(2\epsilon)^{1/2}] \right\}, \\
f_{1/2}(\epsilon) &= \frac{1}{2^{9/2}\pi^3(2+\epsilon)^{9/2}} \Big\{ \tfrac{3}{2}(3 - 32\epsilon - 8\epsilon^2) \sinh^{-1}(\epsilon/2)^{1/2} \\
&\quad - \frac{[\epsilon(2+\epsilon)]^{1/2}}{28}(63 - 693\epsilon - 5670\epsilon^2 - 7410\epsilon^3 - 4488\epsilon^4 - 1448\epsilon^5 - 240\epsilon^6 - 16\epsilon^7) \Big\},
\end{aligned} \tag{21}$$

where $\operatorname{erf}(x)$ is the error function and $F(x) = \exp(-x^2)\int_0^x \exp(t^2)dt$ is Dawson's integral. The expressions for $\eta = 1$ and $2$ are due to Jaffe (1983)[2] and Hernquist (1990). Analytic expressions are also available for other $\eta$-models (e.g. $\eta = \tfrac{2}{3}, \tfrac{4}{3}$), but they are even more complicated.

---

[2] In Jaffe's equation (7) the terms involving $F[(-E)^{1/2}]$ and $F[(-2E)^{1/2}]$ should be multiplied by $-i$.

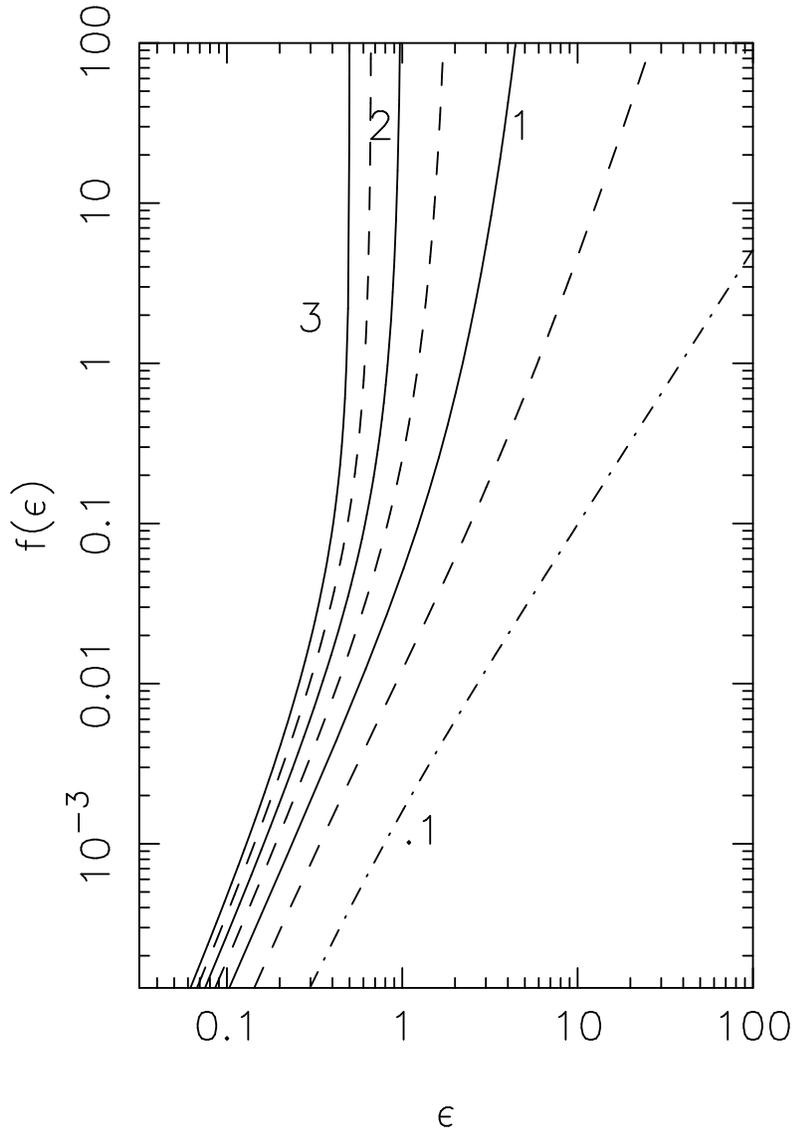

2. Phase-space distribution function (mass per unit phase-space volume) $f(\epsilon)$ as a function of the energy per unit mass $-\epsilon$, for the same values of $\eta$ shown in Figure 1. The curves for $\eta > 1$ diverge at $\epsilon_{\mathbf{max}} = 1/(\eta - 1)$, the depth of the central potential well; for $\eta \leq 1$ the depth of the central potential well is infinite.

Figure 2 shows the DF $f(\epsilon)$ for several $\eta$-models. The DF increases with $\epsilon$ for all $\eta$-models, a result which can be verified analytically and which implies that $\eta$-models are stable (BT).

**Surface brightness**  Since we assume that the mass-to-light ratio $\Upsilon$ is constant, the surface brightness at projected radius $R$ is

$$I(R) = \frac{2}{\Upsilon} \int_R^\infty \frac{\rho(r) r \, dr}{(r^2 - R^2)^{1/2}}. \tag{22}$$

Using equation (3) and the substitution $u = \sin^{-1}(R/r)$, the surface brightness of an $\eta$-model may be written

$$I_\eta(R) = \frac{\eta R^{\eta-2}}{2\pi \Upsilon} \int_0^{\pi/2} \frac{\sin^2 u \, du}{(R + \sin u)^{1+\eta}}. \tag{23}$$

Analytic expressions for $I_\eta(R)$ are available for models with integer values of $\eta$:

$$\begin{aligned}
\Upsilon I_\eta(R) &= \frac{1}{4R} + \frac{1 - (2 - R^2) X(R)}{2\pi(1 - R^2)}, & \eta &= 1, \\
&= \frac{-3 + (2 + R^2) X(R)}{2\pi(1 - R^2)^2}, & \eta &= 2, \\
&= \frac{2 + 13R^2 - (12R^2 + 3R^4) X(R)}{4\pi(1 - R^2)^3}, & \eta &= 3,
\end{aligned} \tag{24}$$

where

$$\begin{aligned}
X(R) &= \frac{1}{(1 - R^2)^{1/2}} \cosh^{-1} \frac{1}{R}, & R &< 1, \\
&= \frac{1}{(R^2 - 1)^{1/2}} \cos^{-1} \frac{1}{R}, & R &> 1.
\end{aligned} \tag{25}$$

The expressions for $\eta = 1, 2$ are due to Jaffe (1983) and Hernquist (1990). The surface brightness can also be written analytically for half-integer $\eta$, using elliptic integrals; however, the expressions are sufficiently complicated that they are not very useful.

The surface brightness profiles satisfy a recurrence relation:

$$I_{\eta+1}(R) = -\frac{R^{\eta-1}}{\eta} \frac{d}{dR} \left[ R^{2-\eta} I_\eta(R) \right]. \tag{26}$$

The asymptotic behavior of the surface brightness is

$$\Upsilon I_\eta(R) \to \frac{\eta}{8 R^3} \qquad \text{as } R \to \infty. \tag{27}$$

At small radii,

$$\begin{aligned}
\Upsilon I_\eta(R) &\to \frac{\eta \Gamma(1 - \frac{1}{2}\eta)^2}{2^{\eta+1} \pi \Gamma(2 - \eta)} R^{\eta-2}, & 0 &< \eta < 2, \\
&\to \frac{1}{\pi} \left[ \ln(2/R) - \tfrac{3}{2} \right], & \eta &= 2, \\
&\to \frac{1}{\pi(\eta - 2)(\eta - 1)}, & 2 &< \eta \leq 3.
\end{aligned} \tag{28}$$

All models with $2 < \eta < 3$ have a core with constant surface brightness even though the volume density diverges as $r^{\eta-3}$: the divergence is weak enough that the emissivity integrated along the line of sight is dominated by radii of order unity rather than by the central cusp.

For any value of $\eta$ the surface brightness may be determined numerically. The numerical evaluation of the integral in equation (23) is straightforward for $R \gtrsim 1$. For $R \ll 1$ numerical integrations are slow to converge because the integrand changes behavior near $u \sim R$ (the integrand is $u^{1-\eta}$ for $R \ll u \ll 1$ but $u^2/R$ for $u \ll R$). A strategy that improves the convergence is to integrate numerically the difference between the integral in (23) and $\int_0^{\pi/2} u^2(R+u)^{-1-\eta}du$, which has the same behavior for $u \ll 1$ but can be integrated analytically. A subroutine that computes $I_\eta(R)$ is available from the authors.

The surface brightness profiles of $\eta$-models are plotted in Figure 3(a). Figure 3(b) compares the surface brightness profiles to de Vaucouleurs' (1948) empirical law, which states that $\log I$ is a linear function of $R^{0.25}$.

**Line-of-sight dispersion** The velocity dispersion along the line of sight at projected radius $R$ is denoted $\sigma_p(R)$ and is given by

$$\begin{aligned}
\sigma_p^2(R) &= \frac{2}{\Upsilon I(R)} \int_R^\infty \frac{\rho(r)\overline{v_r^2}(r) r\, dr}{(r^2-R^2)^{1/2}} \\
&= \frac{2}{\Upsilon I(R)} \int_R^\infty \frac{r\, dr}{(r^2-R^2)^{1/2}} \int_r^\infty \rho(r')\frac{d\Phi}{dr'} dr' \\
&= \frac{2}{\Upsilon I(R)} \int_R^\infty \frac{\rho(r)M(r)}{r^2}(r^2-R^2)^{1/2} dr,
\end{aligned} \qquad (29)$$

where the second line follows from the hydrostatic equilibrium equation (6). Using equations (3) and (4) we have

$$\sigma_{p,\eta}^2(R) = \frac{Y_\eta(R)}{\Upsilon I_\eta(R)} \qquad (30)$$

where

$$\begin{aligned}
Y_\eta(R) &= \frac{\eta}{2\pi} R^{2\eta-3} \int_1^\infty \frac{x^{2\eta-5}(x^2-1)^{1/2}}{(1+Rx)^{1+2\eta}} dx \\
&= \frac{\eta}{2\pi} R^{2\eta-3} \int_0^{\pi/2} \frac{\sin^3 u \cos^2 u\, du}{(R+\sin u)^{1+2\eta}}.
\end{aligned} \qquad (31)$$

The integral for $Y_\eta(R)$ can be evaluated analytically for integer or half-integer $\eta$ (Hernquist 1990 gives the analytic expression for $\eta=2$). However, the integrals are cumbersome. A simpler approach is to use the recursion relation

$$Y_{\eta+1/2}(R) = -\frac{R^{2\eta-2}}{2\eta}\frac{d}{dR}\left[R^{3-2\eta}Y_\eta(R)\right], \qquad (32)$$

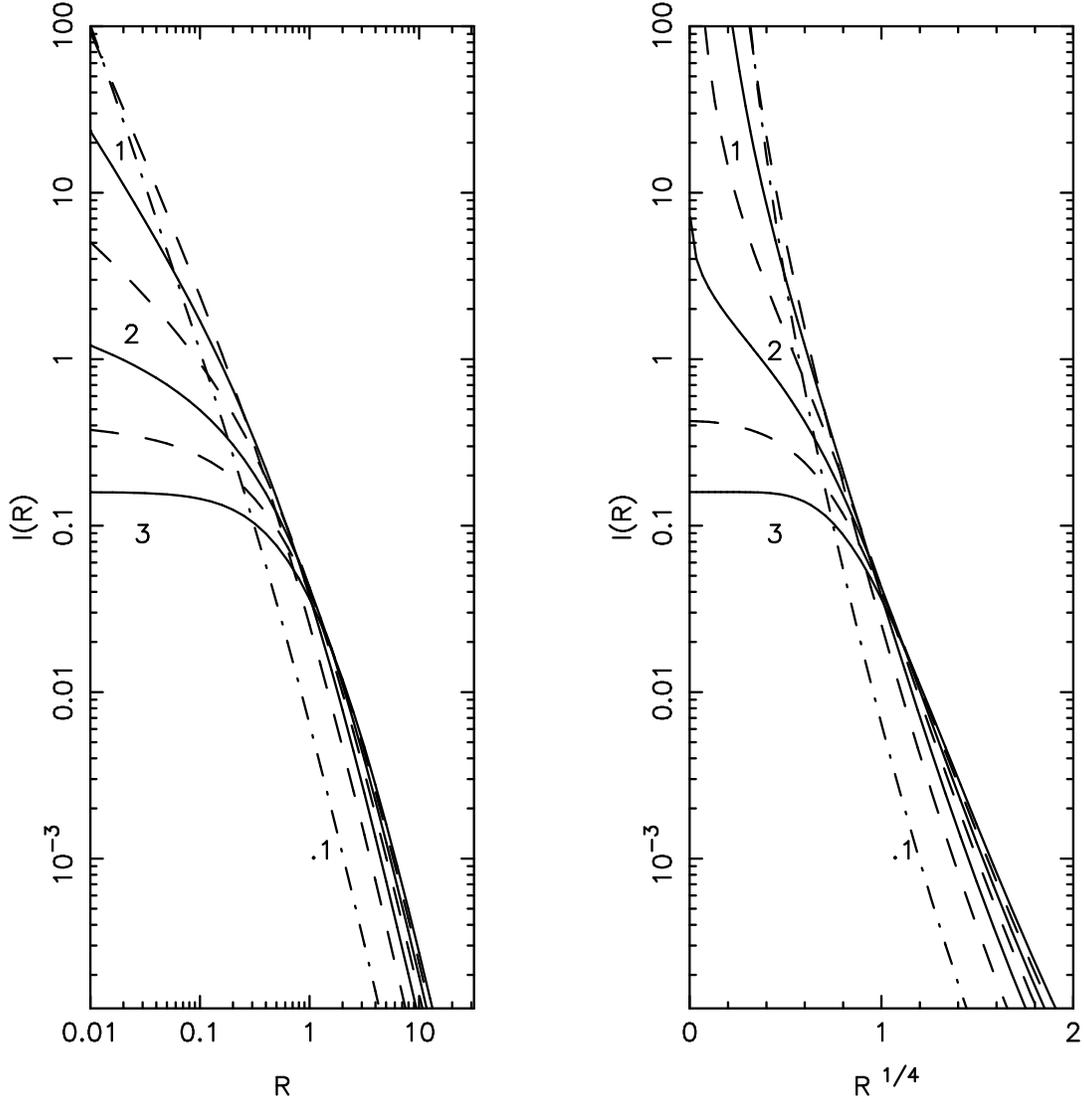

3. Surface brightness of $\eta$-models as a function of radius $R$, normalized so that the total luminosity is unity. The same values of $\eta$ are plotted as in previous figures. (a) Log-log coordinates; (b) Log of the surface brightness plotted against $R^{0.25}$; in these coordinates de Vaucouleurs' surface brightness profile is a straight line.

together with the analytic expression

$$Y_{1/2}(R) = \tfrac{1}{2}R - \frac{1}{\pi} - \frac{1}{8R} + \frac{1}{12\pi R^2} + \frac{(3-4R^2)}{4\pi}X(R), \qquad (33)$$

where $X$ is defined in equation (25).

The asymptotic behavior of the line-of-sight dispersion is given by

$$\sigma^2_{p,\eta}(R) \to \frac{8}{15\pi R} \qquad \text{as } R \to \infty. \qquad (34)$$

The behavior at small radii is most easily given in terms of the variable $Y_\eta$:

$$Y_\eta(R) \to \frac{\eta \Gamma(\frac{3}{2} - \eta)}{8\pi^{1/2}\Gamma(3-\eta)} R^{2\eta-3}, \qquad 0 < \eta < \tfrac{3}{2},$$

$$\to \frac{1}{4\pi}\left[3\ln(2/R) - \tfrac{17}{2}\right], \qquad \eta = \tfrac{3}{2}, \quad (35)$$

$$\to \frac{3}{2\pi(2\eta-1)(2\eta-2)(2\eta-3)}, \qquad \tfrac{3}{2} < \eta \leq 3.$$

From these expressions the behavior of $\sigma_p^2(R)$ can be deduced using equations (28) and (30). For $0 < \eta < 1$, $\sigma_p^2 \propto R^{\eta-1}$ diverges; for $1 < \eta < 2$ it approaches zero at the center (as $R^{\eta-1}$ for $\eta < \tfrac{3}{2}$ and as $R^{2-\eta}$ for $\eta > \tfrac{3}{2}$), and for $\eta > 2$ it is asymptotically constant at the center[3].

The line-of-sight dispersion $\sigma_p(R)$ is plotted in Figure 4 for several values of $\eta$. The subroutine used for these calculations is available from the authors.

**Aperture dispersion** The aperture dispersion $\sigma_a(R)$ is the root-mean-square velocity measured through a circular aperture of radius $R$ centered on the galaxy. Clearly

$$\sigma_a^2(R) = \frac{\int_0^R I(R)\sigma_p^2(R) R\,dR}{\int_0^R I(R) R\,dR}. \qquad (36)$$

The aperture dispersion is only defined for models with $\eta > \tfrac{1}{2}$; otherwise the dispersion measured through any central aperture is infinite. A convenient form for numerical evaluation can be derived from equation (22) and the third line of equations (29):

$$\sigma_a^2(R) = \tfrac{1}{3} \frac{\int_0^\infty \rho(r)M(r)r\,dr - \int_R^\infty \rho(r)M(r)(r^2-R^2)^{3/2}\,dr/r^2}{\int_0^\infty \rho(r)r^2\,dr - \int_R^\infty \rho(r)r(r^2-R^2)^{1/2}\,dr}$$

$$= \tfrac{1}{3} \frac{-W - 4\pi \int_R^\infty \rho(r)M(r)(r^2-R^2)^{3/2}\,dr/r^2}{M - 4\pi \int_R^\infty \rho(r)r(r^2-R^2)^{1/2}\,dr}, \qquad (37)$$

where $M$ and $W$ are the total mass and potential energy (eq. 13). For $\eta$-models with $\eta > \tfrac{1}{2}$ we have

$$\sigma_{a,\eta}^2(R) = \frac{1}{6(2\eta-1)} \frac{1 - 2\eta(2\eta-1) \int_R^\infty \frac{r^{2\eta-5}(r^2-R^2)^{3/2}}{(1+r)^{1+2\eta}}\,dr}{1 - \eta \int_R^\infty \frac{r^{\eta-2}(r^2-R^2)^{1/2}}{(1+r)^{1+\eta}}\,dr}. \qquad (38)$$

At large radii $\sigma_{a,\eta}^2(R) \to 1/[6(2\eta-1)]$. The behavior at small radii is easily deduced from equations (28), (30), (35), and (36). For $0 < \eta < \tfrac{3}{2}$, $\sigma_a^2 \propto R^{\eta-1}$ as $R \to 0$; for $\tfrac{3}{2} < \eta < 2$, $\sigma_a^2 \propto R^{2-\eta}$; for $\eta > 2$, $\sigma_a^2$ is asymptotically constant at the center.

The aperture dispersion is plotted in Figure 5.

---

[3] The leading term of the expression given by Hernquist (1990) for $\sigma_p^2(R)$ as $R \to 0$ is incorrect. The denominator $[8\ln(2/s) - 12]\pi$ should be $4\ln(2/s) - 6$.

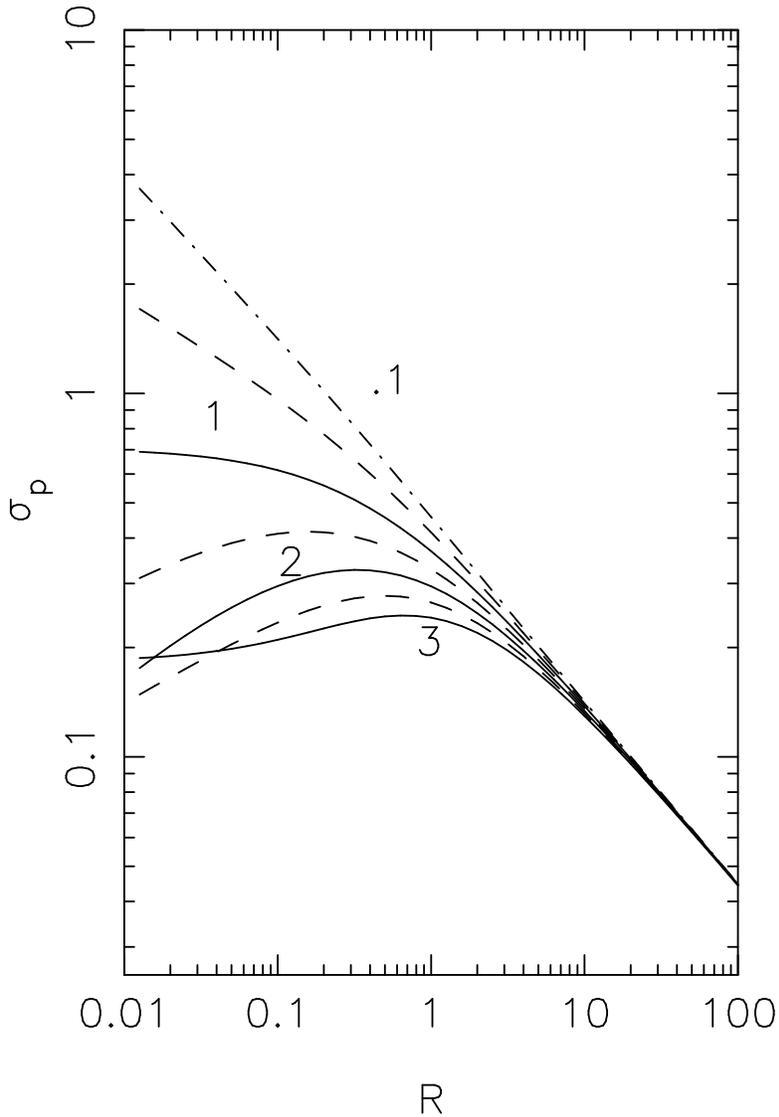

4. Line-of-sight dispersion $\sigma_p$ as a function of radius $R$ for the same models shown in earlier figures.

## 3 Models with a central black hole

We now consider $\eta$-models that contain a non-luminous central object ("black hole") of mass $\mu$. The density $\rho(r)$ (eq. 3) and surface brightness $I(R)$ (eq. 23) are unchanged, and the total mass in stars is still unity. The sole effect of the black hole is to modify the potential $\Phi(r)$ (eq. 5) to

$$\Phi^*(r) \equiv \Phi(r) - \frac{\mu}{r} \tag{39}$$

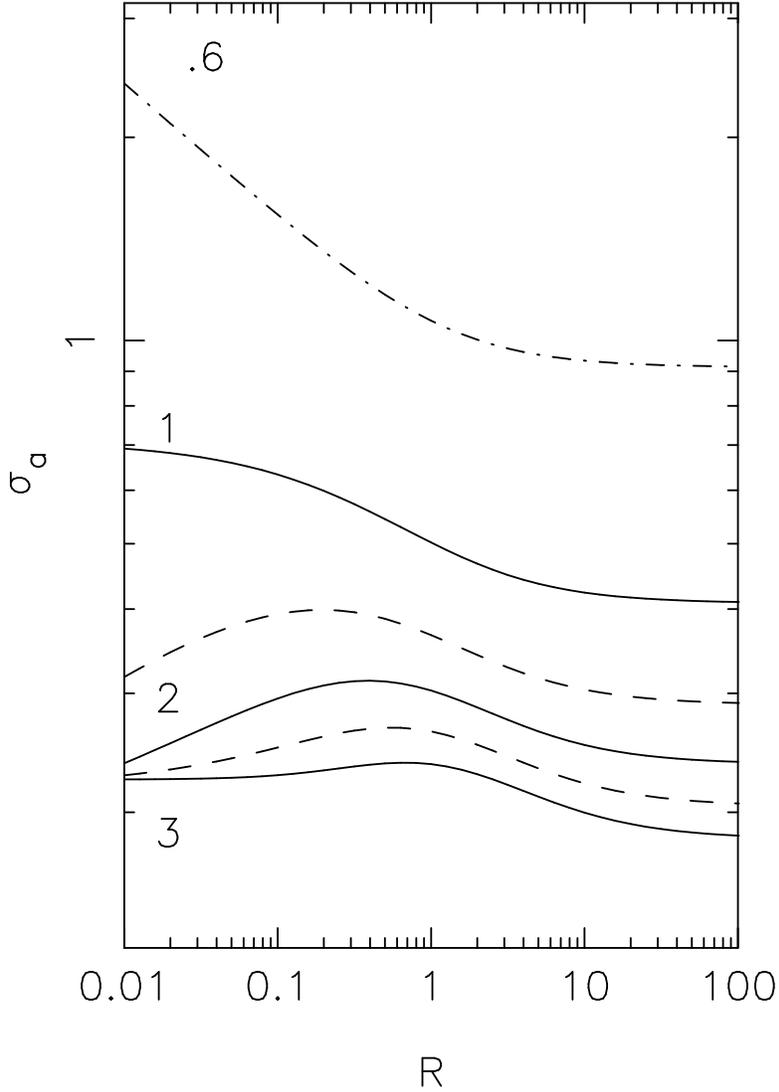

5. Aperture dispersion $\sigma_a$ (the dispersion measured through an aperture of radius $R$ centered on the galaxy) for models with $\eta = 0.6$ (dash-dot line), 1, 2, 3 (solid lines), $\frac{3}{2}$, $\frac{5}{2}$ (dashed lines). The aperture dispersion is infinite for models with $\eta \leq \frac{1}{2}$.

(we denote properties of models with a black hole by the superscript $*$). We restrict ourselves to models with $\eta \leq \frac{5}{2}$ since we show below that models in which the DF depends only on energy are unphysical if $\eta > \frac{5}{2}$.

**Velocity dispersion**      If the velocity-dispersion tensor is isotropic, the radial velocity dispersion is given by

$$\overline{v_r^2}^*(r) = \overline{v_r^2}(r) + \frac{\mu}{\rho(r)} \int_r^\infty \frac{\rho(r)dr}{r^2} \equiv \overline{v_r^2}(r) + \mu s^2(r), \tag{40}$$

where $\overline{v_r^2}(r)$ is given by equations (7) and (8) and

$$s_\eta^2(r) = r^{3-\eta}(1+r)^{1+\eta}\left(\frac{1}{\eta-4} - \frac{4}{\eta-3} + \frac{6}{\eta-2} - \frac{4}{\eta-1} + \frac{1}{\eta}\right) \\ - \frac{r^{-1}(1+r)^5}{\eta-4} + \frac{4(1+r)^4}{\eta-3} - \frac{6r(1+r)^3}{\eta-2} + \frac{4r^2(1+r)^2}{\eta-1} - \frac{r^3(1+r)}{\eta}. \tag{41}$$

The cases $\eta = 1, 2$ are special:

$$\begin{aligned} s_1^2(r) &= \tfrac{1}{3}r^{-1} - \tfrac{1}{3} + \tfrac{4}{3}r + 6r^2 + 4r^3 - 4r^2(1+r)^2 \ln(1+1/r), \\ s_2^2(r) &= \tfrac{1}{2}r^{-1} - \tfrac{3}{2} - 11r - 15r^2 - 6r^3 + 6r(1+r)^3 \ln(1+1/r). \end{aligned} \tag{42}$$

As $r \to 0$,

$$s_\eta^2(r) \to \frac{1}{(4-\eta)r}, \tag{43}$$

which represents the usual $r^{-1}$ divergence in the mean-square velocity expected near a black hole. As $r \to \infty$, $s^2(r) \to \tfrac{1}{5}r^{-1}$ for all $\eta$.

**Distribution function**      The DF can be determined analytically when the energy $\epsilon$ is large (i.e. close to the black hole). We begin by evaluating the integral in equation (16),

$$J(\epsilon) \equiv \int_0^\epsilon \frac{d\rho}{d\psi^*} \frac{d\psi^*}{(\epsilon-\psi^*)^{1/2}} = \int_0^{u(\epsilon)} \frac{d\rho}{du} \frac{du}{[\epsilon-\psi^*(u)]^{1/2}}, \tag{44}$$

where we have changed the integration variable to $u = r^{-1}$ and $u(\epsilon)$ is defined implicitly by $\psi^*[u(\epsilon)] = \epsilon$. We have

$$\begin{aligned} \psi_\eta^* &= -\Phi_\eta + \frac{\mu}{r} = \frac{1}{\eta-1}\left[1-(1+u)^{1-\eta}\right] + \mu u, \qquad \eta \neq 1, \\ &= \ln(1+u) + \mu u, \qquad \eta = 1, \end{aligned} \tag{45}$$

and

$$\frac{d\rho_\eta}{du} = \frac{\eta}{4\pi} \frac{u^3}{(1+u)^{2+\eta}}[4+(3-\eta)u]. \tag{46}$$

When the energy $\epsilon$ is sufficiently large, the integral (44) is dominated by the contribution from the largest values of $u$, so that $\psi_\eta^* \to \mu u$, $d\rho_\eta/du \to [\eta(3-\eta)/4\pi]u^{2-\eta}$, and

$$J_\eta(\epsilon) \to \frac{\eta(3-\eta)}{4\pi} \int_0^{\epsilon/\mu} \frac{u^{2-\eta}du}{(\epsilon-\mu u)^{1/2}} = \frac{\eta\Gamma(4-\eta)}{4\pi^{1/2}\mu^{3-\eta}\Gamma(\tfrac{7}{2}-\eta)}\epsilon^{5/2-\eta}. \tag{47}$$

Thus
$$f^*_\eta(\epsilon) = \frac{1}{2^{3/2}\pi^2}\frac{dJ_\eta}{d\epsilon} = \frac{\eta\Gamma(4-\eta)}{2^{7/2}\pi^{5/2}\mu^{3-\eta}\Gamma(\frac{5}{2}-\eta)}\epsilon^{3/2-\eta}. \tag{48}$$

Since $\Gamma(x) < 0$ for $-1 < x < 0$, models with $\eta > \frac{5}{2}$ are unphysical. There is a simple physical explanation for this constraint: consider a DF consisting of stars at a single energy $E_0$, $f(E) = \delta(E - E_0)$, in the potential of a point mass $\mu$, $\Phi(r) = -\mu/r$. The density is $\rho(r) = 4\pi \int_0^\infty \delta(\frac{1}{2}v^2 - \mu/r - E_0)v^2 dv = 2^{5/2}\pi(E_0 + \mu/r)^{1/2}$. At small radii $\rho(r) \to r^{-1/2}$. Thus, if the DF is isotropic, stars of a given energy always produce a density cusp around a black hole that rises as $r^{-1/2}$ at radii $r \ll \mu/|E_0|$; $\eta$-models with $\eta > \frac{5}{2}$ are not allowed because their density cusps are shallower than this limiting value.

Figure 6 shows the DF for $\eta$-models with several values of the black hole mass $\mu$.

The stability properties of isotropic spherical stellar systems containing a central black hole are not presently known.

**Line-of-sight dispersion**  The mean-square velocity along the line of sight may be written
$$\sigma_p^{2*}(R) = \frac{Y(R) + \mu Z(R)}{\Upsilon I(R)}, \tag{49}$$
where
$$Z_\eta(R) = 2\int_R^\infty \frac{\rho_\eta(r)(r^2 - R^2)^{1/2}}{r^2}dr$$
$$= \frac{\eta}{2\pi}R^{\eta-3}\int_1^\infty \frac{x^{\eta-5}(x^2-1)^{1/2}}{(1+Rx)^{1+\eta}}dx. \tag{50}$$

Comparing to equation (31) we find the convenient identity
$$Z_\eta(R) = 2Y_{\eta/2}(R), \tag{51}$$

so that the program used to compute $Y_\eta$ can be used to compute $Z_\eta$, and analytic expressions for $Z_1$ and $Z_2$ can be found from equations (32) and (33). At small radii,

$$\begin{aligned}\sigma_{p,\eta}^{2*}(R) &\to \frac{2\Gamma(\frac{3}{2}-\frac{1}{2}\eta)^2\mu}{(4-\eta)(2-\eta)\Gamma(1-\frac{1}{2}\eta)^2}R^{-1} \quad 0 < \eta < 2, \\ &\to \frac{\pi\mu}{4R[\ln(2/R)-\frac{3}{2}]}, \quad \eta = 2, \\ &\to \frac{\pi^{1/2}\eta(\eta-1)(\eta-2)\Gamma(\frac{3}{2}-\frac{1}{2}\eta)\mu}{8\Gamma(3-\frac{1}{2}\eta)}R^{\eta-3}, \quad 2 < \eta \leq \frac{5}{2}.\end{aligned} \tag{52}$$

Figure 7 shows the line-of-sight dispersion profile for several $\eta$-models with black holes of varying masses.

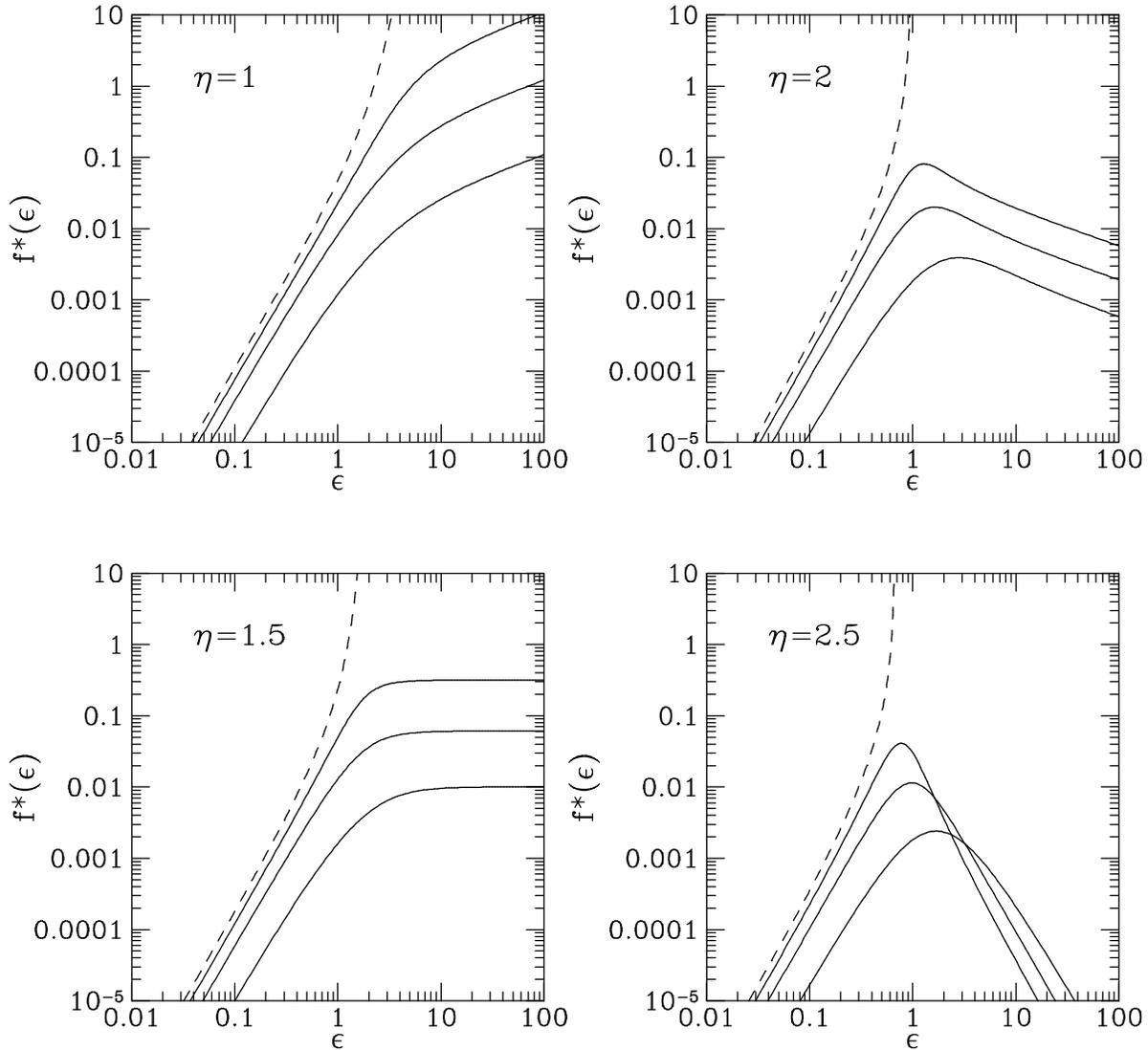

**6.** Phase-space distribution function $f^*$ as a function of energy $\epsilon$ for models with $\eta = 1, \frac{3}{2}, 2,$ and $\frac{5}{2}$, containing central black holes of mass $\mu = 0$ (dashed line), 0.1, 0.3, and 1. The total mass in stars is 1.

## 4 Other "simple" models?

In a strict mathematical sense the $\eta$-models are not particularly simple, since all of the quantities we have investigated—radial velocity dispersion, surface brightness, line-of-sight dispersion, aperture dispersion, and DF—can be derived by quadratures starting with (say) an arbitrary density profile $\rho(r)$. In a practical sense the models are not the simplest possible ones either: except for special values of $\eta$ the quadratures for the surface brightness, line-of-sight dispersion, aperture dispersion and DF are most easily done numerically.

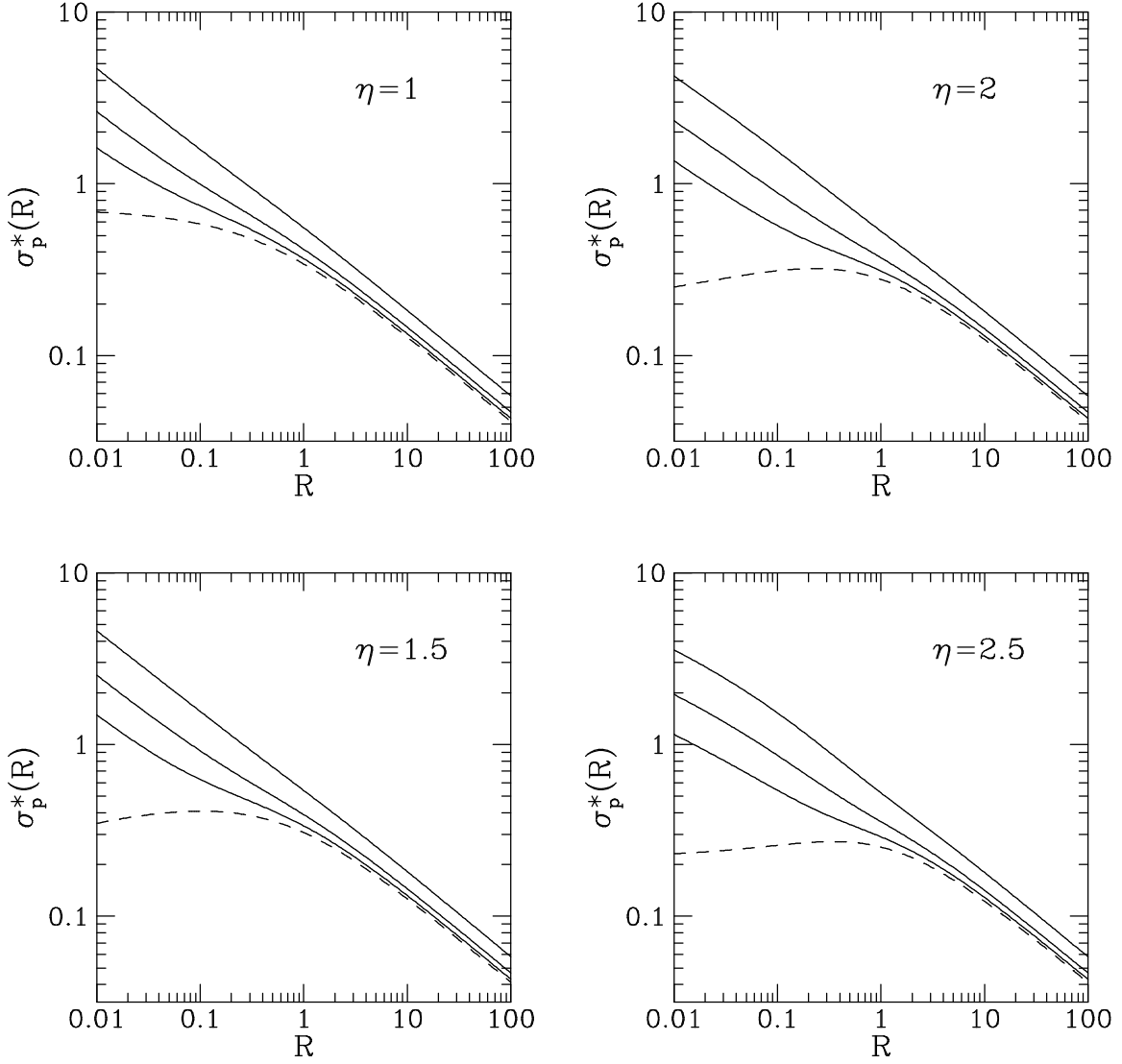

**7.** Line-of-sight dispersion $\sigma_p^*$ as a function of radius $R$ for models with $\eta = 1$, $\frac{3}{2}$, $2$, and $\frac{5}{2}$, containing central black holes of mass $\mu = 0$ (dashed line), $0.1$, $0.3$, and $1$. The total mass in stars is $1$.

Nevertheless the $\eta$-models share several "user-friendly" features: the density $\rho_\eta(r)$, mass $M_\eta(r)$, and potential $\Phi_\eta(r)$ all have simple analytic forms; the mean-square radial velocity $\overline{v_r^2}(r)$ is analytic; and the density can be expressed as a simple function of the potential (which is useful in Eddington's formula [16]). We have failed to find generalizations of or alternatives to the $\eta$-models of comparable simplicity and realism.

It may be useful to describe a chain of argument that leads naturally to the $\eta$-models. In choosing a simple functional form for a galaxy model, the natural place to start is with the potential $\Phi(r)$, since this can be differentiated to yield the mass and density:

$$M(r) = r^2 \frac{d\Phi}{dr}, \qquad \rho(r) = \frac{1}{4\pi r^2}\frac{dM}{dr}. \qquad (53)$$

These operations are simpler when we use the variable $u = 1/r$:

$$M(u) = -\frac{d\Phi}{du}, \qquad \rho(u) = \frac{u^4}{4\pi}\frac{d^2\Phi}{du^2}. \qquad (54)$$

Thus it is natural to look for a simple function $\Phi(u)$ such that $\Phi \to -u$ as $u \to 0$ (which normalizes the total mass to unity); another useful property is that $\Phi(u)$ should be invertible, so that $u$ can be expressed as a function of $\Phi$. A suitable choice is $\Phi(u) = [1-(1+u)^b]/b$, which leads to $\eta$-models with $b = 1 - \eta$. Other choices are possible but the ones we have examined are all less attractive.

Note from equation (54) that the asymptotic behavior $\rho \propto r^{-4}$ as $r \to \infty$ ($\rho \propto u^4$ as $u \to 0$) that is common to all the $\eta$-models is a feature of any functional form that can be expanded in a Taylor series, $\Phi(u) = -u + \sum_{j=2}^{\infty} a_j u^j$ with $a_2 \neq 0$. If the first non-zero coefficient is $a_k$, the density at large radii falls as $r^{-(2+k)}$.

## 5 Discussion

The models that we have described have many limitations: they are spherical, whereas most elliptical galaxies and spiral bulges are probably triaxial; the DF is isotropic in velocity space, an assumption that has no strong justification; and their common surface brightness profile at large radii ($I \propto R^{-3}$) does not adequately represent the diverse behavior of the outer envelopes of real galaxies. Despite these shortcomings, the $\eta$-models provide useful illustrations of the kinematic and photometric behavior that can be present in the central parts of galaxies. They also offer simple "strawman" models for comparison with observations of galaxies that exhibit photometric power-law cusps in their centers.

### 5.1 Classification of central structure

As an illustration, let us imagine that a galaxy is observed to have a surface brightness profile $I(R)$ which is $\propto R^{-\gamma}$ at small radii, falling more steeply at radii larger than some characteristic radius that we shall call the "break radius". We also assume that the galaxy is spherical and its DF is isotropic in velocity space. We distinguish "true" radii, which are measured in three dimensions from the center of the galaxy, from "projected" radii, which are the components of the true radius vector that are normal to the line of sight. Our results imply that three types of central structure are possible, not just for $\eta$-models but for any galaxy with these general features:

(I) *flat core* structure ($\gamma = 0$, corresponding to $3 \geq \eta > 2$): A density cusp as steep as $\rho \propto r^{-1}$ may be present in this type, but there is no cusp in the surface brightness, which is dominated by stars whose true radii are of order the break radius. When observed with limited resolution, some galaxies in this class may appear to have small but non-zero values of $\gamma$ (cf. Fig. 3). If there is no central black hole, the depth of the central potential well is finite; the line-of-sight dispersion is asymptotically constant as $R \to 0$ and is dominated by stars at true radii comparable to the break radius. If a black hole is present, the mean-square line-of-sight velocity grows, at a rate between $\sigma_p^2 \propto R^{-1}$ and $R^{-0.5}$ depending on $\eta$; models with $\eta > \frac{5}{2}$ are unphysical in this case. Hernquist's model ($\eta = 2$) marks the boundary between types I and II.

(II) *weak cusp* structure ($0 < \gamma < 1$, $2 > \eta > 1$): In this type $\eta = 2 - \gamma$; thus the density cusp is $\rho \propto r^{-\gamma-1}$. The surface brightness is dominated by stars whose true radii are comparable to the projected radius. If no black hole is present the central potential has finite depth and the mean-square line-of-sight velocity approaches zero near the center, $\sigma_p^2 \propto R^\gamma$ for $\gamma < \frac{1}{2}$ and $\sigma_p^2 \propto R^{1-\gamma}$ for $\gamma > \frac{1}{2}$. If a black hole is present the mean-square velocity rises as $\sigma_p^2 \propto R^{-1}$. Jaffe's model ($\gamma = 1$) marks the boundary between Types II and III.

(III) *strong cusp* structure ($1 < \gamma < 2$, $1 \geq \eta > 0$): In this type, as in the previous one, $\eta = 2 - \gamma$ and the density cusp is $\rho \propto r^{-\gamma-1}$. The central potential well has infinite depth, and the mean-square line-of-sight velocity diverges near the center, $\sigma_p^2 \propto R^{1-\gamma}$ with no black hole and $\propto R^{-1}$ if a black hole is present.

### 5.2 Measuring mass-to-light ratios

The traditional method of determining the mass-to-light ratio of spherical stellar systems with flat cores is known as core fitting or King's method (King 1966, Richstone and Tremaine 1986). The determination is based on the formula for the mass-to-light ratio

$$\Upsilon = k \frac{9\sigma_p^2(R=0)}{2\pi G I(R=0) R_{\rm hb}}, \tag{55}$$

where $G$ is the gravitational constant (set to unity in earlier sections), $R_{\rm hb}$ is the half-brightness radius defined by $I(R_{\rm hb}) = \frac{1}{2}I(0)$, and $k$ is nearly unity for a wide variety of stellar systems (and hence is set to unity in the usual applications of the method).

Core fitting clearly is not an adequate approach to measuring the mass-to-light ratios of most $\eta$-models, or of any galaxy with similar central structure. To illustrate the difficulties, we evaluate the constant $k$ in equation (55) as a function of the parameter $\eta$. For $\eta=3$, $k = 1.048$; $k$ rises to 1.091 at $\eta = 2.64$, then sinks to 0.900 by $\eta = 2.32$. Thus for $2.32 < \eta \leq 3$ the core fitting formula (55) with $k = 1$ is accurate to better than 10%. However, for values of $\eta$ outside this limited range the accuracy of the core fitting formula plummets: at $\eta = 2.2$ we have $k = 0.50$, at $\eta = 2.1$ we have $k = 0.05$, and for $\eta \leq 2$ the formula fails completely since the central surface brightness diverges.

Other methods must replace core fitting for models with $\eta \lesssim 2.3$:

(i) The *local* method is based on the surface brightness $I(R)$ and line-of-sight dispersion $\sigma_p(R)$ measured at a single radius; dimensional analysis then yields the formula

$$\Upsilon = h \frac{\sigma_p^2(R)}{GRI(R)}, \tag{56}$$

where $h(\eta, R)$ is a dimensionless constant that is plotted in Figure 8. When $\eta < \frac{3}{2}$ and $R$ is much smaller than the break radius, $h$ is given by the analytic formula

$$h = \frac{\eta(2-\eta)\Gamma(1-\frac{1}{2}\eta)^4}{2^{2\eta-1}\pi^{3/2}\Gamma(2-\eta)\Gamma(\frac{3}{2}-\eta)}, \quad 0 < \eta < \frac{3}{2}. \tag{57}$$

The local method is appropriate when the radius $R$ is large compared to the spatial resolution of the observations.

(ii) The *aperture* method is based on the aperture dispersion $\sigma_a(R)$ and the luminosity inside an aperture, $S(R) \equiv 2\pi \int_0^R I(R')R'dR'$:

$$\Upsilon = g \frac{R\sigma_a^2(R)}{GS(R)}, \tag{58}$$

where $g(\eta, R)$ is a dimensionless constant that is plotted in Figure 8. The aperture method fails for $\eta < \frac{1}{2}$ because the aperture dispersion diverges. When $\frac{1}{2} < \eta < \frac{3}{2}$ and $R$ is much smaller than the break radius, $g$ is given by the analytic formula

$$g = \frac{(2-\eta)(2\eta-1)\Gamma(1-\frac{1}{2}\eta)^4}{2^{2\eta-2}\pi^{1/2}\eta\Gamma(2-\eta)\Gamma(\frac{3}{2}-\eta)}, \quad \frac{1}{2} < \eta < \frac{3}{2}. \tag{59}$$

The aperture method is appropriate when the radius $R$ is close to the limiting spatial resolution.

### 5.3 The formation process

Our discussion so far has focused exclusively on the range of possible equilibrium models, without addressing issues of galaxy formation. Different formation processes may favor particular values of $\eta$. Several examples of such constraints are known for the case where a central black hole is present. If the black hole grows slowly, arguments based on adiabatic invariance imply $\gamma = \frac{1}{2}$ (Peebles 1972, Young 1980, BT), although in this case the velocity-dispersion tensor is somewhat anisotropic. If a steady-state distribution of stars around the black hole has been established by two-body relaxation, we expect $\gamma = \frac{3}{4}$ for stars of equal mass (Bahcall and Wolf 1976, BT). If the formation process leaves no stars bound to the black hole, we expect $\gamma = 0$, $\eta = \frac{5}{2}$ (Peebles 1972, BT).

Programs to compute the surface brightness, distribution function, line-of-sight and aperture dispersion of $\eta$-models are available by electronic mail from the authors. ST

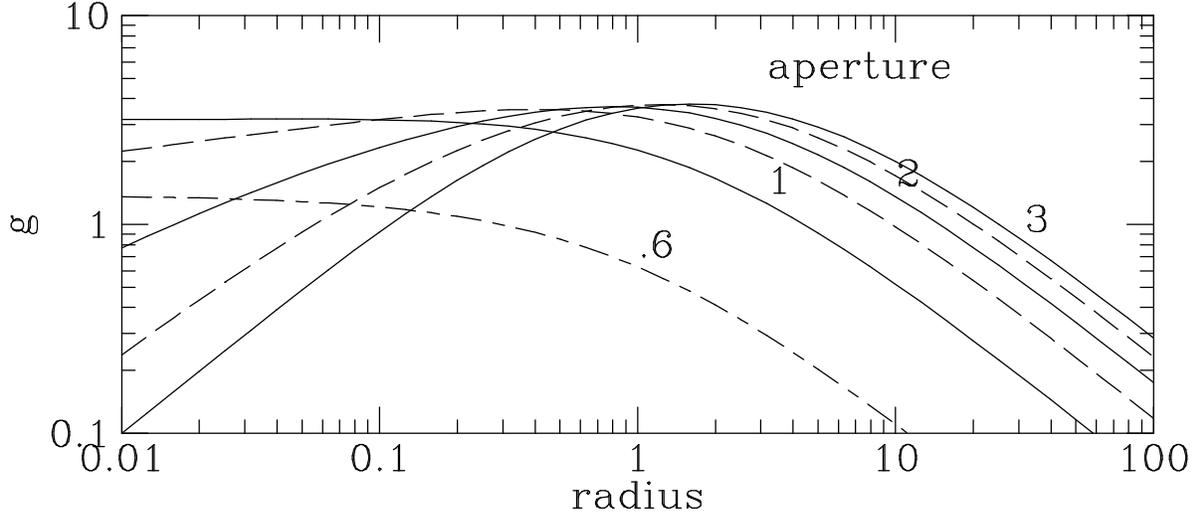

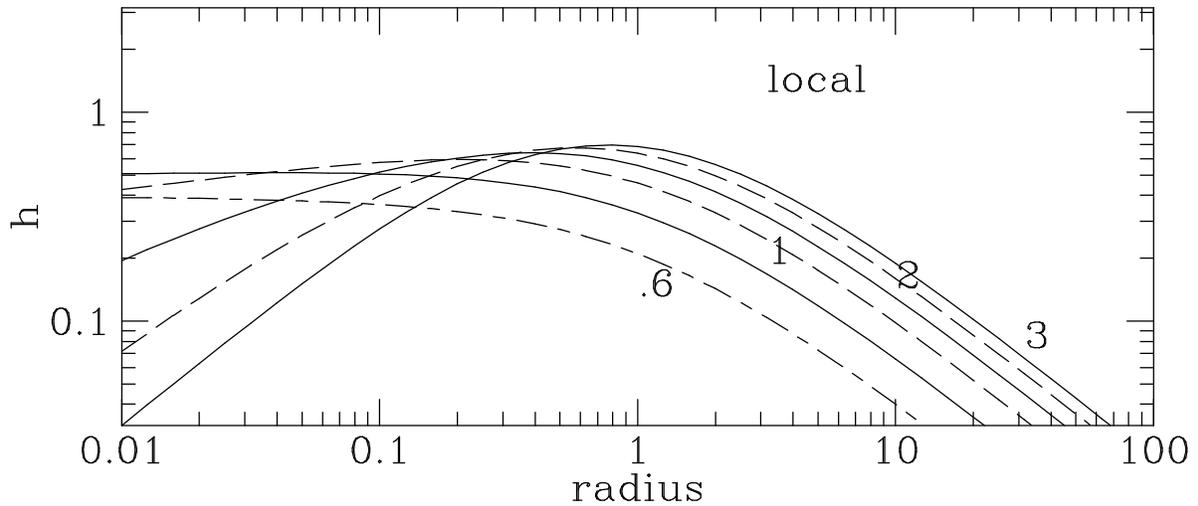

8. The dimensionless quantities $g$ and $h$ (eqs. 56 and 58) that determine the mass-to-light ratio from measurements of the velocity dispersion and surface brightness, plotted as a function of radius $R$ for models with $\eta = 0.6$ (short dash-long dash curve), 1, 2, 3 (solid curves), $\frac{3}{2}$, $\frac{5}{2}$ (unlabelled dashed curves).

thanks Steve Balbus and especially Prasenjit Saha for discussions and advice. DOR thanks the Institute for Advanced Study for hospitality during a portion of this work. Support for this work was provided by a research grant from NSERC, and by NASA through General Observer Grants 2600 and 3912 and grant number HF-1029.01-92A from the Space Telescope Science Institute, which is operated by the Association of Universities for Research in Astronomy, Inc. under NASA contract NAS5-26555.